\documentclass[prl,twocolumn,amsfonts]{revtex4}
\usepackage{graphicx}
\usepackage{bm}

\newlength{\figwidth}
\begin{document}

\title{Direct imaging of periodic sub-wavelength patterns of total atomic density}
\author{Alexei Tonyushkin}
\email{alexei@nyu.edu}
\author{Tycho Sleator}
\affiliation{New York University, Department of Physics,
4 Washington Place, New York, New York 10003}
%\date{}
\begin{abstract}
Interference fringes of total atomic density with period $\lambda /4$ and $\lambda /2$ 
for optical wavelength $\lambda$, 
have been produced in de Broglie atom interferometer and directly imaged by means of 
an ``optical mask'' technique. 
The imaging technique allowed us to observe sub-wavelength periodic patterns 
with a resolution of $\lambda /16$. 
The quantum dynamics near the interference times as a function of the recoil phase and pulse areas
has been investigated.
\end{abstract}
%\pacs{}
\maketitle

An interferometer based on the interaction of a pair of off-resonant standing wave pulses 
(made from laser fields with wavelength $\lambda$) with a cold gas of Rb atoms \cite{Cahn97} 
has been shown to be capable of precision measurement of the atomic recoil frequency, $\hbar/m$ 
and inertial forces such as gravity~\cite{Snadden98}-\cite{Lenef97Gustavson97}.  

If the standing-wave pulses in this interferometer are separated by time $T$,
theory predicts that at times $t \sim [(n+1)/n] T$ for positive integer $n$, fringe patterns 
of the total atomic density of period 
$\lambda/2n$ should appear~\cite{Dubetsky94}. Such fringes are the manifestation of matter-wave 
diffraction~\cite{Dubetsky84,Batelaan97}. 
Evidence of these $\lambda/2n$ period gratings has been observed indirectly in Ref.~\cite{STKS02} 
by use a backscattering technique in a heterodyne arrangement, 
but no direct observation of small-period fringes has been made until now.  

To observe such small period structures directly, we have
developed a real-time imaging technique for $^{85}$Rb atoms \cite{Turlapov03}. 
This ``optical mask''~\cite{Abfalterer97Keller98, Johnson98} technique, 
was applied to atoms of $^{85}$Rb initially prepared in the ground hyperfine level $F=3$.  
An ``optical mask'' is a standing wave (SW) resonant to 
the $F=3\!\rightarrow\! F^\prime =3$ transition (5S$_{1/2} \!\rightarrow\,$5P$_{3/2}$, 
$\lambda=780$ nm). For a pulse of sufficient
intensity and duration, atoms not located at the nodes of the SW are
optically pumped into the F=2 level.
For imaging the density at a particular time, a
``detection sequence'' is applied, consisting of an optical mask followed by a
traveling wave pulse tuned to the closed transition
$F=3 \! \rightarrow\! F^\prime$=4.
Atoms left unpumped at the nodes following the mask pulse are
counted by observation of fluorescence in the traveling wave.  The
fluorescent signal is proportional to the density at the nodes
just before the application of the imaging mask.
To map out the density as a function of position, the initial density profile is
reproduced and the measurement repeated with various locations of
the detection mask node within the mask period of $\lambda/2$.

In this work we applied the ``optical mask'' technique to directly image the fringe structures in an atom
 interferometer consisting of two off-resonant SW pulses separated by a time $T$, 
each consisting of two linearly polarized traveling waves blue detuned from the 
$F=3\!\rightarrow\! F^\prime=4$ transition.  
Each SW pulse acts as a phase grating for the atoms. 
These phase gratings do not change the internal state of an atom, but only alter its center of mass. 
%(atomic momentum). Each momentum state 
A given initial atomic momentum state is split into a superposition of momentum states differing by twice the photon momentum.
This superposition results in a evolving fringe pattern that washes out due to 
a spread in the initial momentum distribution of the atoms.  
%the spread in momenta of the incoming waves. 
The second pulse results in removal of the initial momentum dependence of the gratings at various times (referred to as {\it echo's}).  The times of the echoes can be predicted by a classical analysis, but the size and shape of the fringes can only be understood in terms the quantum behavior.  

At times close to the echo time $t=T(n+1)/n$, one expects to find a periodicity in the atomic density of $\lambda/2n$ because the minimum difference between the
momenta of interfering states is $n\hbar {\bf q}$, where ${\bf q=k_1-k_2}=2{\bf k}$ is a grating vector for a SW consisting of two traveling waves (${\bf k_1, k_2}$). 

A simple interference signal calculation (assuming Raman-Nath regime), which only takes into 
the account the lowest spatial harmonic of the density grating is presented in Ref.~\cite{STKS02}. 
We can extend this result to include all harmonics by writing the resulting density as a Fourier series:
% based on Ref[9]:
\begin{equation}
\label{rhoF}
\rho({\bf x},t)=\sum_{N_{2}=-\infty}^{+\infty}
\rho_{N_{2}}(t)\: e^{i N_{2}{\bf q\cdot x}}\,,
\end{equation}
where the $N_{2}$th Fourier harmonic is built up by interfering
matter waves whose momenta differ by $N_{2}$ recoils after the second SW. 
% TS
Because of Doppler dephasing, gratings occur only near echo times 
\begin{equation}
\label{echotime}
 t^{echo}_{\bf N}= T(1 - N_1/N_2)\,,
\end{equation}
where ${\bf N} \equiv \{N_{1}, N_{2}\}$, $N_1$ is an integer, and $N_1/N_2 < 0$.  
%
% at this point we give new calculations: 
Near a given echo time $t^{echo}_{\bf N}$,
the amplitude of the $N_{2}$th Fourier harmonic of the density for $\Delta t=t-t^{echo}_{\bf N}$ 
is given by
\begin{eqnarray}
\label{AverhoN2}
\rho_{N_{2}} (t_{\bf N}^{echo} +\Delta t) =  e^{-(N_{2} q u \Delta t/2)^{2}}J_{N_{1}}[2\theta_{1}\sin(N_{2} \omega_{q} \Delta t)] 
\nonumber \\ \times J_{N_{2}-N_{1}}[2\theta_{2}\sin\varphi_{2}^{rec}(t^{echo}_{\bf N} + \Delta t)] \,,
\end{eqnarray}
where the recoil phases: $ \varphi^{rec}_{1}(t)=\omega_{q} \left[N_{2}(t-T)+ N_1 T\right]$,  
$\varphi^{rec}_{2}(t)=\omega_{q} N_{2}(t-T)$. 
Here $\theta _i $ are pulse areas, $\theta_i \sim \Omega \tau_i$, 
where $\Omega$ is a  two photon Rabi frequency and $\tau_i$ are the pulse durations. 
The recoil frequency $ \omega _q =\hbar {\bf q} ^2/2m_{atom} $, and $u$ is the spread in initial velocities.

A density pattern of period $\lambda /2 n$ at an echo point can be obtained by letting $-N_1/N_2 = m/n$, where positive integers $m$ and $n$ are expressed in lowest terms, and the sum in Eq.~(\ref{rhoF}) only includes values of $N_2$ for which both $N_1$ and $N_2$ are integers.  We get the corresponding echo time $ t_{n,m} ^{echo}=T (n+m)/n$.

In our experiments we consider only the primary echoes for which $m=1$. From  Eqs.~(\ref{rhoF})-(\ref{AverhoN2}), one
can get the general expression for the spatial profile of the total atomic density grating 
of period $\lambda /2n$ formed at the echo time $ t_{n} ^{echo}=T (n+1)/n $ :
\begin{eqnarray}
\label{genrho}
\rho (x, t_n ^{echo} + \Delta t)=  \sum _{N=-\infty}^{\infty } 
(-1)^N e^{-(n N q u \Delta t/2)^2} %\hspace{0.9in} 
\nonumber \\ \times
J_N\left[ 2 \theta _1 \sin \left( n N \omega _q \Delta t \right) \right] 
\\ \times
J_{ (n+1) N}\left\{ 2 \theta _2 \sin \left[ N \omega _q (T+n \Delta t) \right] \right\} \, e^{i n N q x} \nonumber \, .
\end{eqnarray}
Because of the exponential term in the sum, for small periodicity ($\lambda/2n$ with $n\gg 1$) 
only the first harmonic $N=1$ contributes significantly to the signal.
However for $n$ small (for example $\lambda/2$ periodicity at time $2T$) higher harmonics also contribute 
to the signal and define its spatial profile. 

Further look at Eq.~(\ref{genrho}) shows periodic behavior of the atomic density as a function of 
$T$ with the period $\pi/\omega_q\simeq 32\,\mu$s. This behavior was first observed in~\cite{Cahn97}.
It is interesting to look at the time dependence of the density at the echo time around $2T$ 
for $\Delta t\to 0$. From Eq.~(\ref{genrho}) we get $\rho (2T+\Delta t) \sim \Delta t$ 
at all times except when $\omega_{q} T=m\pi$ (for integer $m$), in which case $\rho (2T+\Delta t)\sim {\Delta t}^3$,
which shows the significance of the recoil phase. Also there are no interference fringes at exactly the echo times, 
only an atomic phase modulation exists at these times. This is in a contrast to the situation for {\em absorptive} 
gratings~\cite{Weitz04, Turlapov05}, in which the interference fringes have maximum amplitude exactly at the echo times. 
The above expressions can also be used to estimate the temperature
of the trapped atoms ($\sim u$), since the duration of the signal is $\sim 1/(qu)$.

\begin{figure}[htb!]
\begin{center}
\includegraphics[width=\figwidth]{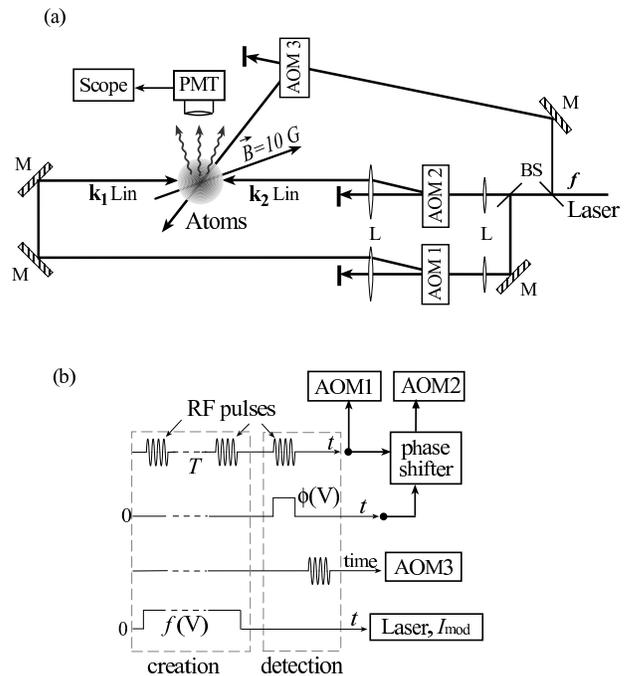}
\end{center}
\caption{Experimental setup for the imaging the structures by optical mask:
a) Optical beam diagram (M - mirror, BS - beam splitter, AOM - acousto-optic
modulator, L - lens);
b) Pulse diagram which shows rf pulses switching 
optical $k_1, k_2$ beams and the detecting traveling wave. Also shown on the diagram are the 
pulses that change the phase between $k_1$ and $k_2$ and the frequency of the laser.}
\label{ExpSetup}
\end{figure}

In the experiment $^{85}$Rb atoms are prepared from a vapor in a MOT.
The experiment is done in the time domain with pulsed laser fields and repeated
every 50 ms. The experimental setup and the time diagram of the experiment are shown in Fig.~\ref{ExpSetup}.
%%%%%%%%%%%%%%%%%%%%%%%%%%%%%%%%%%%%%%%%%%%%%%%%%%%%%%%%%%
%%%%%%%%%%%%%%%%%%%%%%%%%%%%%%%%%%%%%%%%%%%%%%%%%%%%%%%%%%%%

At times $t=0$ and $t=T$ two off-resonant standing-wave pulses (SW1,
SW2) are applied with typical pulse durations of 200--480 ns.  The pulses
are blue-shifted from the closest transition $5S_{1/2}(F=3)\!\rightarrow \!
5P_{3/2}(F^{\prime}=4)$ of $^{85}$Rb by a detuning ($\Delta$) that varied from 30 MHz to 
105 MHz ($\simeq$ 5 to 18 excited state linewidths correspondingly). With this detuning, less than 5 \% of 
the atoms spontaneously emit during each pulse.
The SW pulses are composed of two counter-propagating traveling waves
${\bf k}_{1}$ and ${\bf k}_{2}$ with the intensity in each beam 
200 mW/cm$^{2}$, which are switched on and off independently by
two acousto-optic modulators (AOMs), driven by a common 220 MHz rf oscillator.

The density grating of the atomic cloud is probed around specific echo times  
by a detection sequence consisting of an optical mask
SW pulse followed by a weak traveling wave (aligned in the horizontal plane at an angle $45^\circ$ to the 
mask beams) for measuring the fluorescence. The optical mask standing wave is produced by the same optics that produce the 
off-resonant standing waves.  To shift the frequency of the SWs 
between different phases of the experiment, we modulated the laser 
current.  Each of the traveling waves making up the optical mask has 
horizontal linear polarization and an intensity of 40 to 120 mW/cm$^{2}$. 
The nodes of the optical mask are shifted relative to the nodes of 
the off-resonant SWs by a phase change of the rf feeding one of the AOMs. 
The read-out field on the cycling $F=3\!\rightarrow\! F^\prime=4$ transition 
is pulsed by a separate AOM. The use of the same optics for both sets of pulses allowed us to 
minimize the phase drift during a single repetition of the experiment. 
In our experiment, both off-resonant SWs used to
create the pattern have the same pulse areas $\theta_1 =\theta_2 =\theta$. However 
one can mimic $\theta_2 \ne \theta_1 $ by varying recoil phase ($\sim T$), since the density
function [Eq.~(\ref{genrho})] is periodic in both the recoil phase and the pulse area.

\begin{figure}[htb!]
\begin{center}
\includegraphics[width=\figwidth]{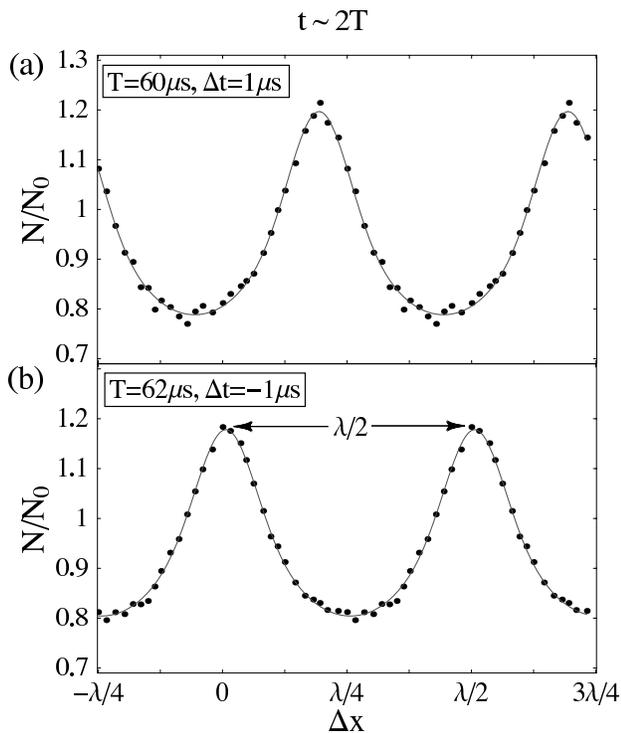}
\end{center}
\caption{Total atomic density as a function of a phase shift of the optical mask SW in the vicinity of 
an echo time $2T$.  The curve is a model function fit to the data.  The actual data were taken for
$\Delta x  \in \{-\lambda /4$, $\lambda /4\}$ and then repeated on the graph with the period $\lambda /2$:
a)$ T=60 \mu$s, $\Delta t=1\mu$s, $\theta=4.5$, $\eta=0.94$, $V=22\%$; b)$ T=62 \mu$s, $\Delta t=-1\mu$s,
$\theta=4.6$, $\eta=0.94$, $V=20\%$;. The data are normalized to the uniform
atomic density (no off-resonant pulses applied prior to the imaging sequence).}
\label{echosignal2T}
\end{figure}

In Fig.~\ref{echosignal2T} we plot the spatial structures with the period $\lambda /2$, observed 
at the time $\sim 2T$. 
%%%%%%%%%%%%%%%%%%%%%%%%%%%%%%%%%%%%%%%%%%%%%%%%%%%%%%%%%%%%%%%%
%%%%%%%%%%%%%%%%%%%%%%%%%%%%%%%%%%%%%%%%%%%%%%%%%%%%%%%%%%%%%%%%%%%%%%%%%%%%%%%%%%%%%%%%%%%%%%%%%%%%%%
%and: $\Delta t=1\mu$s [Fig.~\ref{echosignal2T}(a)];$\Delta t=-1\mu$s, T=62$\,\mu$s [Fig.~\ref{echosignal2T}(b)]. 
To maximize visibilities of the fringes the data were taken with slightly different values of the recoil phase 
($\sim T$) for the two graphs with pulse area fixed. 
There is a phase shift of $\pi$ between the two fringe patterns taken for opposite $\Delta t$ 
(up to a systematic shift in the phase calibration over the several day interval between two data sets). 
This phase shift is a characteristic feature of atomic density patterns created with optical phase gratings and 
it is predicted by the theory. The data on the graphs are normalized to 
the uniform atomic density distribution obtained when no off-resonant SWs are applied 
prior to the imaging sequence. The atomic loss is represented by the parameter $\eta$
obtained from the model fit to the data. In the ideal case of no atomic loss $\eta=1$, in practice it 
is 6\% -- 13\% less than that. The background due to stray photons (no atoms present in the MOT) is subtracted.

All the data are fitted by a model function (curve in Fig.~\ref{echosignal2T}), that is
the convolution of Eq.~(\ref{genrho}) with a Gaussian transmission function representing the effect of the optical 
mask~\cite{Turlapov03}.
The model function allows one to determine important parameters, such as the 
pulse area $\theta$, the width of the optical mask $\sigma$, and an
atomic loss $\eta$. 
The temperature of the atoms can be deduced given that (from the fit) the Doppler broadening term is 
$1/qu \simeq 1.5 \, \mu$s, corresponding to a temperature of $\sim 15\, \mu$K. 

The particular choice of the pulse area $\theta$ and the separation between the pulses $T$ 
was chosen to maximize the visibility 
of the fringes, which is defined by $V\equiv (\rho_{max}-\rho_{min})/(\rho_{max}+\rho_{min})$.
For the data in Fig.~\ref{echosignal2T} the signal visibility is $V=22\%$ (a) and $V=20\%$ (b).
The atomic density visibility, however, is $V_0 \sim 33\%$ (determined by effectively deconvolving the optical mask
transmission function from the signal). 
%%%%%%%%%%%%%%%%%%%%%%%%%%%%%%% 2D

\begin{figure}[htb!]
\begin{center}
\includegraphics[width=\figwidth]{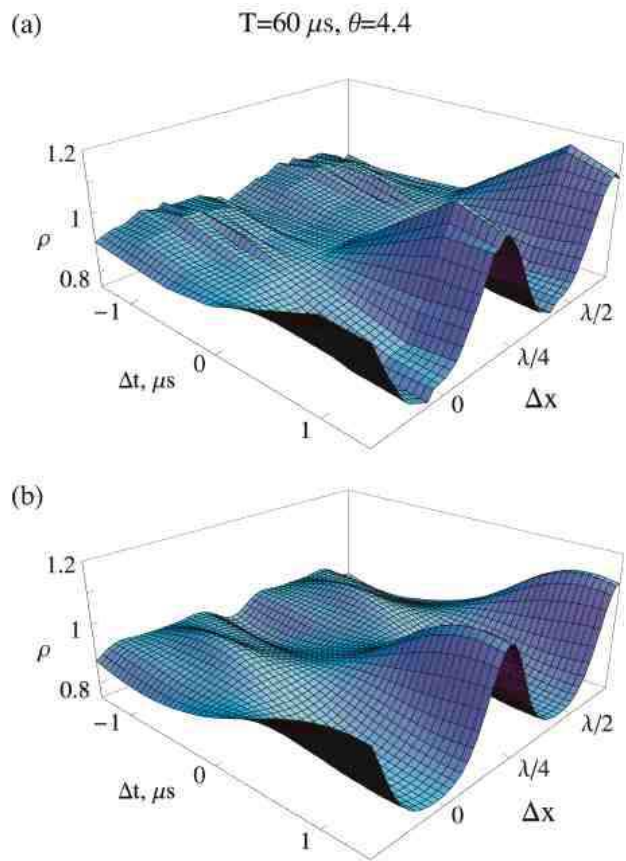}
\end{center}
\caption{Dynamics of the interference fringes of the atomic density in a time
$\Delta t=t-2T$ in the vicinity of an echo time 2T for T=60 $\mu$s: a) Interpolation of the data; 
b) Model function plot for $\theta=$4.4.
 The data are normalized to the uniform atomic density.}
\label{3Decho0505}
\end{figure}

By combining several cross-sections of spatial profiles with period $\lambda /2$, each
 taken at a different value of $\Delta t$, one can directly view the dynamics of the interference fringes, 
reconstructing effectively a ``two-dimensional'' 
profile of the atomic density, where one coordinate is a time $\Delta t$ and the other is a spatial
phase $\Delta x$. 
In Fig.~\ref{3Decho0505} we present such plots.
Fig.~\ref{3Decho0505}(a) represents the interpolation of the data, taken for 
%%%%%%%%%%%%%%%%%%%%%%%%%%%%%%%%%%%%%%%%%%%%%%%%%%%%%%%%%%
%%%%%%%%%%%%%%%%%%%%%%%%%%%%%%%%%%%%%%%%%%%%%%%%%%%%%%%%%%%%%%%%%%%%%%%%%%%%%%%%%%%%%%%%%%%%%%%%%%%%%%
$\Delta x  \in \{-\lambda /4$, $\lambda /4\}$ and repeated with the period $\lambda /2$. Part (b)
shows the corresponding theoretical function for the atomic density [Eq.~(\ref{genrho})] convolved 
with the optical mask transmission function. The data show asymmetric behavior of the fringe pattern
around time $t=2T$ because of particular choice of the recoil phase $\omega_{q} T$ close to $2\pi$ at a fixed pulse area.

We have also directly observed the spatial profile of an atomic fringe pattern with sub-wavelength
period (here $\lambda /4$).
In this experiment the imaging sequence was applied at $t\sim 3T/2$ and the intensities of the 
pulses were the same as for the data in Fig.~\ref{echosignal2T}. The duration of the pulses were adjusted to
maximize the contrast of the signal. 

\begin{figure}[htb!]
\begin{center}
\includegraphics[width=\figwidth]{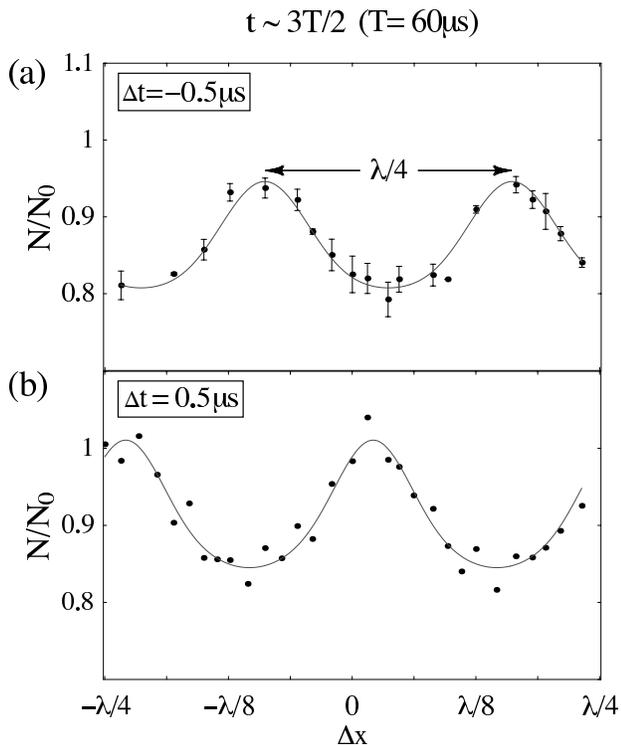}
\end{center}
\caption{Total atomic density as a function of a phase shift of the optical mask SW in the vicinity of 
an echo time $3 T/2$, the curve is a model function fit to the data: a) $\Delta t =-0.5 \,\mu$s, 
$\theta=3.2$, $\eta=0.87$, $V=9\%$, error bars
represents statistics from several runs; b) $\Delta t =0.5 \,\mu$s, $\theta=4.2$, $\eta=0.91$, $V=12\%$;
 The data are normalized to the uniform atomic density.}
\label{echo3_2T0513} 
\end{figure}

The resulting atomic density patterns are shown in
Fig.~\ref{echo3_2T0513}. 
%%%%%%%%%%%%%%%%%%%%%%%%%%%% Sub-wave
%%%%%%%%%%%%%%%%%%%%%%%%%%%%%%%%%%%%%%%%%%%%%%%%%%%%%%%%%%%%%%%%%%%%%%%%%%%%%%%%%%%%%%%%%%%%%%%%%%%%%%
The observed sub-wavelength period structures have lower visibility than for the density of period  $\lambda /2$, 
due to the fact that interfering trajectories with high number of the recoils have lower amplitudes. 
Also for the $\lambda /2n$ (here $ n=2$) period patterns, maximum visibility occurs at 
smaller $\Delta t$ than for $n=1$ due to more quickly decaying term $e^{-(n N q u {\Delta t}/2)^2}$ in 
the expression for the total atomic density Eq.~(\ref{genrho}). 
The maximum visibility of the signal, $V=12\%$, is obtained for $\Delta t =0.5\,\mu$s, T=60$\mu$s 
[see Fig.~\ref{echo3_2T0513}(b)]. For Fig.~\ref{echo3_2T0513}(a) and (b) visibilities of the signals are $V=9\%$ 
$V=12\%$. By taking into the account the instrumental resolution of the mask, 
we compute the visibilities in atomic density to be $V_0 \approx 20\%$ and $V_0 \approx 22\%$ correspondingly.

The fringe patterns at times $t=3T/2 - \Delta t$ and $t=3T/2 + \Delta t$ also have a relative phase shift of $\pi$ 
(between (a) and (b) parts of Figs.~\ref{echo3_2T0513}).
The atomic density with sub-wavelength period ($\lambda /4$ and smaller)
has slightly different dynamics around $\Delta t=0$ than the $\lambda /2$ period fringes.
For the recoil phase $\omega_q T$ is equal to an integer multiple of $\pi$,
 for $\Delta t \to 0 $ the density function behaves as $\sim \Delta t^4$. 
This is a manifestation of quantum behavior of the atomic center of mass, 
%(not predicted classically)
and any spatial modulation of the signal is extremely difficult to observe for such conditions. 

It is worth pointing out, in contrast to the experiments with the absorptive light masks~\cite{Turlapov05}, 
the atom losses from the phase gratings used in this experiment are very small for the data presented.
To date, we have produced patterns with period $\ge \lambda /4$ although smaller periodicities could be obtained
in a similar fashion up to the ultimate resolution of the optical mask imaging technique of $\lambda /16$. 

To conclude, we have directly observed and investigated atomic gratings of total atomic density with period $\lambda /2n$ for
$n\ge 2$ in an atom interferometer.

%\begin{acknowledgments}
We are grateful to A.~Turlapov for useful discussions. 
%\end{acknowledgments}

%%%%%%%%%%%%%%%%%%%%%%%%%%%%%%%%%%%%%%%%%%%%%%%%%%%%%%%%%%%%

\end{document}